\makeatletter
\def\cl@chapter{}
\makeatother

\RequirePackage{fix-cm}

\documentclass[twocolumn,epjc3]{svjour3}  

\RequirePackage[T1]{fontenc}
\RequirePackage{graphicx}
\RequirePackage{mathptmx}      
\RequirePackage{flushend}
\RequirePackage[numbers,sort&compress]{natbib}
\RequirePackage[colorlinks,citecolor=blue,urlcolor=blue,linkcolor=blue]{hyperref}

\journalname{Submitted to Eur. Phys. J. C}

\usepackage{booktabs}

\usepackage[bottom]{footmisc}
\usepackage[utf8]{inputenc}
\usepackage{amsmath,array}
\usepackage{graphicx}  
\usepackage{hyperref}
\usepackage[all]{hypcap}  
\usepackage{dcolumn}                 
\usepackage{bm}                          
\usepackage{epstopdf}
\usepackage{soul}
\usepackage{footnote}
\usepackage{textgreek}
\usepackage{xparse}
\usepackage[capitalise]{cleveref} 
\usepackage{subfigure}
\usepackage{pifont}
\usepackage{color}
\usepackage{multirow} 
\usepackage{upgreek}
\usepackage{siunitx}

\newcommand{\dd}{\mathrm{d}}
\newcommand{\transpose}{^{\mathsf{T}}}

\usepackage{siunitx}
  \DeclareSIUnit\gauss{G}
  \DeclareSIUnit\cps{cps}
  \DeclareSIUnit\year{yr}
\AtBeginDocument{\usepackage{booktabs}} 

\makeatletter
\renewcommand\paragraph{\@startsection{paragraph}{4}{\z@}%
	{1.5ex\@plus 0ex \@minus -.2ex}%
	{1.5ex \@plus .2ex}%
	{\normalfont\normalsize\textit}}
\makeatother

\begin{document}



\institute{%
Institute for Technical Physics~(ITEP), Karlsruhe Institute of Technology~(KIT), Hermann-von-Helmholtz-Platz 1, 76344 Eggenstein-Leopoldshafen, Germany\label{a}
\and Institute for Nuclear Physics~(IKP), Karlsruhe Institute of Technology~(KIT), Hermann-von-Helmholtz-Platz 1, 76344 Eggenstein-Leopoldshafen, Germany\label{b}
\and Technische Universit\"{a}t M\"{u}nchen, James-Franck-Str. 1, 85748 Garching, Germany\label{c}
\and IRFU (DPhP \& APC), CEA, Universit\'{e} Paris-Saclay, 91191 Gif-sur-Yvette, France \label{d}
\and Institute of Experimental Particle Physics~(ETP), Karlsruhe Institute of Technology~(KIT), Wolfgang-Gaede-Str. 1, 76131 Karlsruhe, Germany\label{e}
\and Laboratory for Nuclear Science, Massachusetts Institute of Technology, 77 Massachusetts Ave, Cambridge, MA 02139, USA\label{f}
\and Institute for Data Processing and Electronics~(IPE), Karlsruhe Institute of Technology~(KIT), Hermann-von-Helmholtz-Platz 1, 76344 Eggenstein-Leopoldshafen, Germany\label{g}
\and Institute for Nuclear Research of Russian Academy of Sciences, 60th October Anniversary Prospect 7a, 117312 Moscow, Russia\label{h}
\and Max-Planck-Institut f\"{u}r Kernphysik, Saupfercheckweg 1, 69117 Heidelberg, Germany\label{i}
\and Max-Planck-Institut f\"{u}r Physik, F\"{o}hringer Ring 6, 80805 M\"{u}nchen, Germany\label{j}
\and Department of Physics and Astronomy, University of North Carolina, Chapel Hill, NC 27599, USA\label{k}
\and Triangle Universities Nuclear Laboratory, Durham, NC 27708, USA\label{l}
\and Department of Physics, Faculty of Mathematics and Natural Sciences, University of Wuppertal, Gau\ss{}str. 20, 42119 Wuppertal, Germany\label{m}
\and Departamento de Qu\'{i}mica F\'{i}sica Aplicada, Universidad Autonoma de Madrid, Campus de Cantoblanco, 28049 Madrid, Spain\label{n}
\and Center for Experimental Nuclear Physics and Astrophysics, and Dept.~of Physics, University of Washington, Seattle, WA 98195, USA\label{o}
\and Nuclear Physics Institute of the CAS, v. v. i., CZ-250 68 \v{R}e\v{z}, Czech Republic\label{p}
\and Institut f\"{u}r Kernphysik, Westf\"{a}lische Wilhelms-Universit\"{a}t M\"{u}nster, Wilhelm-Klemm-Str. 9, 48149 M\"{u}nster, Germany\label{q}
\and Helmholtz-Institut f\"{u}r Strahlen- und Kernphysik, Rheinische Friedrich-Wilhelms-Universit\"{a}t Bonn, Nussallee 14-16, 53115 Bonn, Germany\label{r}
\and Department of Physics, Carnegie Mellon University, Pittsburgh, PA 15213, USA\label{s}
\and Nuclear Science Division, Lawrence Berkeley National Laboratory, Berkeley, CA 94720, USA\label{t}
\and University of Applied Sciences~(HFD)~Fulda, Leipziger Str.~123, 36037 Fulda, Germany\label{u}
\and Department of Physics, Case Western Reserve University, Cleveland, OH 44106, USA\label{v}
\and Institut f\"{u}r Physik, Johannes-Gutenberg-Universit\"{a}t Mainz, 55099 Mainz, Germany\label{w}
\and Institut f\"{u}r Physik, Humboldt-Universit\"{a}t zu Berlin, Newtonstr. 15, 12489 Berlin, Germany\label{x}
\and Project, Process, and Quality Management~(PPQ), Karlsruhe Institute of Technology~(KIT), Hermann-von-Helmholtz-Platz 1, 76344 Eggenstein-Leopoldshafen, Germany    \label{y}
}

\author{%
M.~Aker\thanksref{a,b}
\and K.~Altenm\"{u}ller\thanksref{c,d}
\and M.~Arenz\thanksref{r}
\and W.-J.~Baek\thanksref{e}
\and J.~Barrett\thanksref{f}
\and A.~Beglarian\thanksref{g}
\and J.~Behrens\thanksref{b,e}
\and A.~Berlev\thanksref{h}
\and U.~Besserer\thanksref{a,b}
\and K.~Blaum\thanksref{i}
% ...
}

\thankstext{mertensmail}{mail: mertens@mpp.mpg.de}
\thankstext{deceased}{Deceased}
\thankstext{schloessermail}{mail: magnus.schloesser@kit.edu}
\thankstext{also1}{Also affiliated with Oak Ridge National Laboratory, Oak Ridge, TN 37831, USA}


\title{First operation of the KATRIN experiment with tritium}

\maketitle

\begin{abstract}
    The determination of the neutrino mass is one of the major challenges in astroparticle physics today. Direct neutrino mass experiments, based solely on the kinematics of $\upbeta$-decay, provide a largely model-independent probe to the neutrino mass scale. The Karlsruhe Tritium Neutrino (KATRIN) experiment is designed to directly measure the effective electron antineutrino mass with a sensitivity of \SI{0.2}{\electronvolt} (\SI{90}{\percent} CL). In this work we report on the first operation of KATRIN with tritium which took place in 2018. 
    During this commissioning phase of the tritium circulation system, excellent agreement of the theoretical prediction with the recorded spectra was found and stable conditions over a time period of 13 days could be established. These results are an essential prerequisite for the subsequent neutrino mass measurements with KATRIN in 2019.
\end{abstract}

\section{Introduction}
The neutrino mass is non-zero as proven by the discovery of neutrino oscillations~\cite{Fukuda:1998mi, Ahmad:2002jz, Eguchi:2002dm}; however, it is at least five orders of magnitude smaller than the mass of other fermions of the Standard Model of elementary particle physics. The experimental determination of the absolute neutrino mass scale is essential to reveal the origin of neutrino masses and to understand their roles in the evolution of structure in the universe. Cosmological observations~\cite{Aghanim:2018eyx} and the determination of the half-life of neutrinoless double $\upbeta$-decay \cite{Dolinski2019} provide powerful means to probe the neutrino mass. However, they rely on model assumptions. The most model-independent approach is based exclusively on the kinematics of single $\upbeta$-decays~\cite{Otten:2008zz, Drex13}. 

The most advanced one among the direct neutrino mass experiments is the Karlsruhe Tritium Neutrino (KATRIN) experiment. KATRIN is designed to measure the effective electron antineutrino mass $m_{\bar{\upnu}_e}$ with a sensitivity of \SI{0.2}{\electronvolt} (\SI{90}{\percent} CL)~\cite{Angrik:2005ep}. KATRIN's measurement principle is based on a precise determination of the shape of the tritium beta decay ($\text{T} \rightarrow {^{3}\mathrm{He}^{+}} + e^{-} + \bar{\upnu}_e$) spectrum close to its endpoint at about $E_0 = \SI{18.6}{\kilo\electronvolt}$. A non-zero neutrino mass distorts the shape of the $\upbeta$-electron spectrum in the close vicinity of this endpoint. 

A major challenge in detecting this minuscule spectral distortion arises because a fraction of only $10^{-9}$ of all decays generate an electron in the last \SI{40}{\electronvolt}, where the signal of the neutrino mass is maximal. Experimental requirements to overcome this challenge are 1) the operation of a high-activity tritium source, 2) an eV-scale energy resolution, 3) a low background rate, and 4) a well-understood theoretical description of the spectral shape. The optimal isotope is tritium, which features a rather short half-life of 12.3 years, a well-known theoretical representation, and a low endpoint of \SI{18.6}{\kilo\electronvolt}.

\begin{figure*}[]
\centering
		\includegraphics{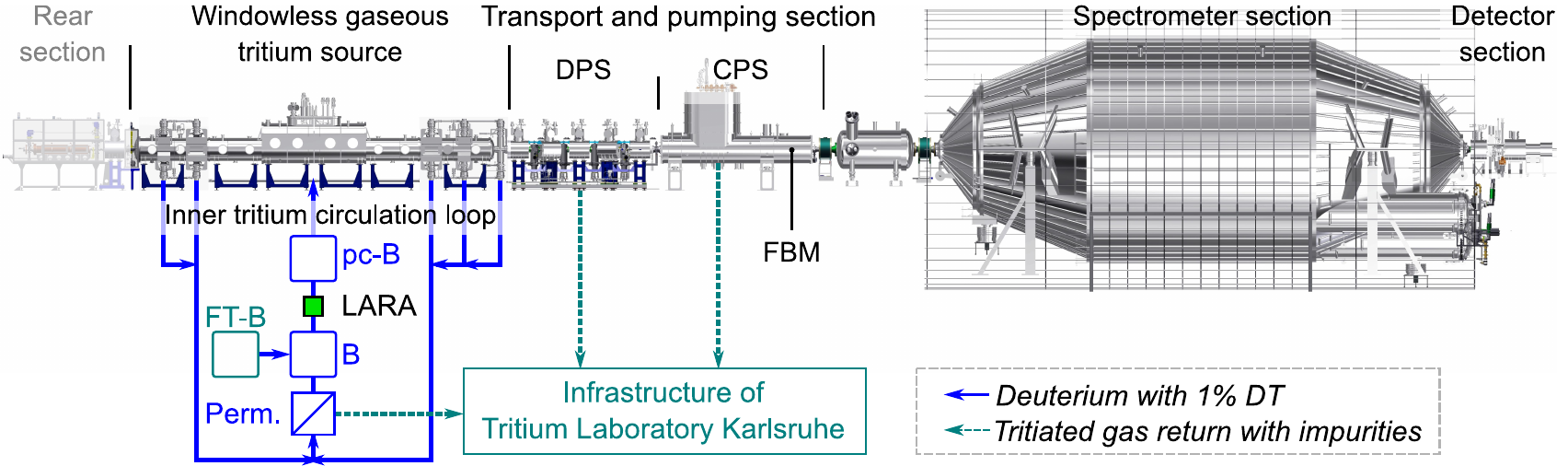}
		\label{Fig:KATRINsetup}
        \caption{The experimental setup of the 70-m-long KATRIN beamline with a conceptual sketch of the tritium loop in the configuration during the First Tritium campaign. FT-B: Gas buffer with pre-defined gas mixture: $\SI{1}{\percent}$ DT in $\mathrm{D_2}$. pc-B, B: (pressure-controlled) buffer vessels. LARA: compositional monitoring by Laser Raman spectroscopy. Perm.: Permeator for hydrogen purification. FBM: Forward Beam Monitor. DPS: Differential Pumping Section. CPS: Cryogenic Pumping Section. The rear section (grayed out) was not used during the FT campaign.}
\end{figure*}

The 70-m long KATRIN beamline, depicted in figure~\ref{Fig:KATRINsetup}, combines a high-luminosity (\SI{e11}{decays\per\second}) gaseous, molecular tritium (T$_2$) source with a high-resolution spectrometer using Magnetic Adiabatic Collimation in an Electrostatic (MAC-E) filter~\cite{Lobashev:1985mu, PICARD1992345}. Tritium decays in the central, 10-m long part of the Windowless Gaseous Tritium Source (WGTS) cryostat~\cite{HeizmanWGTS2017}. The $\upbeta$-electrons are magnetically guided by a system of super-conducting solenoids through the transport and pumping sections towards the spectrometer section. The transport and pumping section reduces the flux of neutral tritium molecules by at least 14 orders of magnitude and rejects tritium ions before they can reach the spectrometer section producing background. The large main spectrometer acts as a MAC-E filter, transmitting only electrons with a kinetic energy $E$ above the retarding energy $qU$ (where $q$ is the elementary charge and $U$ is the retarding voltage of the spectrometer). At the end of the beamline a segmented Si-detector with 148 pixels (focal plane detector, FPD~\cite{Amsbaugh201540, Wall201473}) counts the number of transmitted electrons as a function of retarding voltages of the main spectrometer. The shape of the integral $\upbeta$-electron spectrum is obtained by counting at a pre-defined set of different retarding voltages.

In 2016, all components of the beamline were integrated for the first time and successfully commissioned with electrons and ions created at the rear-end of the KATRIN setup. The alignment of all magnets and the blocking of positive ions were demonstrated~\cite{Arenz:2018kma}. In 2017, the system was further tested with a gaseous and a condensed $^{83\mathrm{m}}$Kr source, demonstrating the excellent spectroscopic performance of the MAC-E filter technology~\cite{Altenmuller:2019ddl} and verifying the calibration of the high-precision high voltage system at the ppm-level~\cite{Arenz:2018ymp}. The success of these two campaigns was the prerequisite for proceeding with the first tritium injection into the WGTS. The analysis of the data obtained in this First Tritium (FT) campaign is the subject of this work.

\section{The First Tritium campaign}
In the FT campaign, the WGTS was mostly operated at the nominal column density of $\rho d = 4.46\cdot10^{17}$ molecules/cm$^2$, however at \SI{0.5}{\percent} of the nominal activity. This safety limitation was achieved by mixing traces of tritium with pure deuterium~\cite{Priester2019, Schloesser2019}. Figure~\ref{Fig:KATRINsetup} illustrates the technical implementation of the gas inlet into the WGTS. A pre-defined gas mixture (\SI{1}{\percent} DT in $\mathrm{D_2}$; $\approx20\,\mathrm{bar}\,\ell$ which corresponds to \SI{9.6}{\tera\becquerel}) was prepared before the campaign in the Tritium Laboratory Karlsruhe (TLK). This gas mixture was circulated through the WGTS via the main tritium loop~\cite{Priester:2015bfa}. The injection into the beamline was regulated by a pressure-controlled buffer vessel. The return gas from the WGTS turbo-molecular pumps was filtered by a palladium-silver membrane (permeator) which is only permeable to hydrogen isotopes. The main part of the flow was reinjected into the WGTS, while a small fraction of the flow including all impurities was continuously sent back to the TLK infrastructure for re-processing. In order to maintain a constant gas flow, an equivalent small amount of DT-D$_2$ gas mixture was injected into the loop from the buffer vessel. At all times the gas composition was monitored by a Laser Raman spectroscopy system~\cite{Sturm2010, Schloesser2013}. The gas circulation was maintained without interruption for the 13 days, which was the complete duration of the FT campaign.

An important difference of the experimental setup during the FT campaign compared to the final experimental configuration of KATRIN concerns the rear section of the beamline: In the full completed experimental configuration the rear section is equipped with an electron gun for calibration purposes and a gold-plated rear wall at the end of the WTGS beam tube for defining and biasing the source electric potential, see figure~\ref{Fig:KATRINsetup}. During the FT campaign this section was not available. The WGTS was instead terminated by a stainless steel gate valve.

\begin{figure}[]
\centering
		\includegraphics{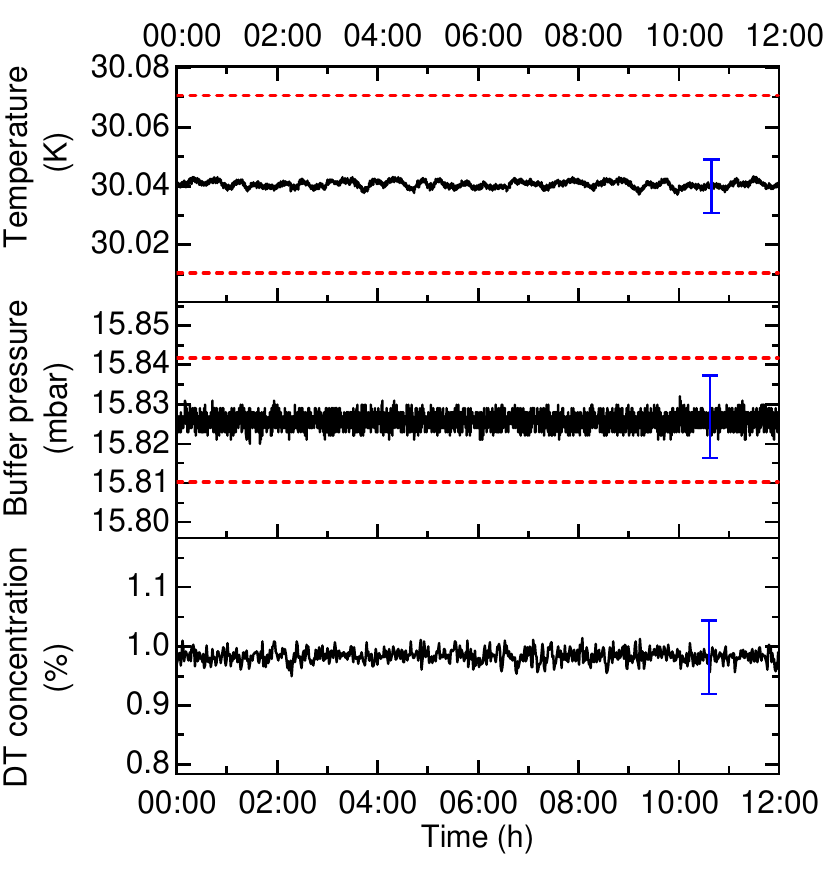}
		\label{Fig:sourcestability}
        \caption{Time stability of beam tube temperature (top panel), buffer vessel pressure (middle panel), and DT concentration (bottom panel) over a time period of 12 hours. The dashed red lines indicate the range of the KATRIN specifications. As KATRIN in its final configuration will operate with 200 times more tritium in the form of T$_2$, no design value for small amounts of DT is defined. The blue error bars indicate the systematic uncertainty on the absolute value of the respective parameters.}
\end{figure}

A key aspect of the FT campaign was to demonstrate a source stability at the \SI{0.1}{\percent} level on the time scale of hours. Important slow-control parameters determining the rate of tritium decays in the source volume are: 1) the beam tube temperature, 2) the buffer vessel pressure, and 3) the isotopic purity~\cite{Babutzka:2012xd}. Figure~\ref{Fig:sourcestability} displays the stability of these parameters over a representative time period of \SI{12}{\hour}. Both the temperature and pressure show time variations on the 10-ppm level, which is more than one order of magnitude better than the required limit. The measurement of the DT concentration fluctuates at the level of \SI{1}{\percent}, which arises from the low amount of DT available for the Laser-Raman measurement and the resulting large (relative) statistical uncertainty. 

In addition, the stability of source activity relies on a constant conductivity of the inlet capillaries. This condition was fulfilled during the FT campaign, where the measured throughput was fully governed by the buffer-vessel pressure. When operating at higher tritium purity, the conductivity can be affected by the production of secondary impurities, which can freeze onto the capillary and beam tube surfaces. 

In order to constantly monitor the source activity a Forward Beam Monitor (FBM) is installed in the KATRIN beamline downstream of the cryogenic pumping section, see figure~\ref{Fig:KATRINsetup}. It is situated outside the magnetic flux tube mapped on the detector and continuously monitors the rate of $\upbeta$-electrons with two silicon \textit{p-i-n} diodes~\cite{Ellinger:2017pek}. Another means of measuring the source activity is by monitoring intermittently the rate of $\upbeta$-electrons with the focal plane detector itself, while keeping the main spectrometer voltage at a fixed and low retarding potential. For a retarding energy of $qU=E_0-\SI{1000}{\electronvolt}$ the $\upbeta$-electron rate of $\SI{20.87}{\kilo\cps}$ in 60 second time-bins was demonstrated to be stable on the \SI{0.1}{\percent} level over a duration of five hours. This stability is fully consistent with Poissonian rate fluctuations.

Beyond these successful stability measurements, a major goal of the FT campaign was to record tritium $\upbeta$-electron spectra. The objectives of these spectral measurements were 1) to compare various analysis strategies, 2) to test the spectrum calculation software, and 3) to demonstrate the stability of the fit parameters in the analyses. 

For the FT measurement, the statistical sensitivity to the neutrino mass was only approximately \SI{6}{\electronvolt} (\SI{90}{\percent} CL), which is much larger than the current bound of \SI{2}{\electronvolt} at (\SI{95}{\percent} CL)~\cite{Tanabashi:2018oca} from the Mainz~\cite{Kraus:2004zw} and Troitsk~\cite{PhysRevD.84.112003} measurements. Consequently, the neutrino mass was fixed to zero in the FT analysis; the endpoint $E^{\text{fit}}_0$ was used instead as a proxy to evaluate the analysis results.

\section{Spectral measurement} \label{sec:specmeas}
\begin{figure}[]
\centering
		\includegraphics[width=84mm]{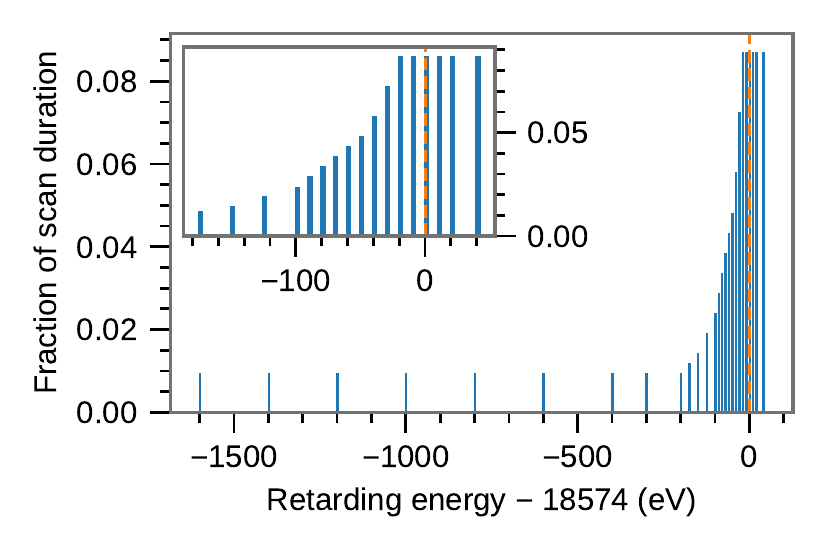}
		\label{Fig:MTD}
        \caption{Typical measurement time distribution for a tritium spectrum scan of 3 hours.
        The inset shows in detail the region closer to the endpoint of $E_0(\mathrm{DT})$ whose approximate value is marked by the dashed line. A scan with fine voltage steps is performed close to the endpoint, adjusting the measurement time at each retarding potential to obtain approximately equal statistics at each setting. Additional wider-spaced measurement points further away from the endpoint and above the endpoint allows the inference of the signal and background rates. }
\end{figure}

KATRIN obtains the integral $\upbeta$-electron spectrum by sequentially applying different retarding energies $qU_i$ to the main spectrometer and counting the number of transmitted $\upbeta$-electrons $N(qU_i)$ with the focal plane detector. The choices of the retarding potentials and the measurement time at a given $qU_i$ are optimized in order to obtain the maximal sensitivity to the parameter of interest and robustness against systematic uncertainties. Figure~\ref{Fig:MTD} shows the measurement time distribution used during FT data taking. 

The spectrum was measured at 30 different retarding potentials in the range of $E_0 - \SI{1600}{\electronvolt} \leq qU_i \leq E_0 + \SI{30}{\electronvolt}$. This interval is significantly larger than the nominal interval for neutrino mass measurements, which typically only extends down to tens of electronvolts below the endpoint. This enlarged interval is a unique feature of the FT campaign, which was technically feasible due to the reduced activity, and hence reduced counting rate at the focal plane detector. The larger interval allowed one to 1) obtain significant statistics to test the treatment of systematic uncertainties (which typically increase further away from the endpoint), 2) gain confidence in our calculation of the spectrum over a wider interval, 3) perform a search for sterile neutrinos in the $200 - \SI{1000}{eV}$ mass range, which is the subject of a separate publication. 

The sequence in which the retarding potentials are applied is alternating between increasing and decreasing voltage (\textit{up-scans} and  \textit{down-scans}). This choice optimizes the averaging of possible drifts of slow-control parameters (for example, the beam tube temperature, high-voltage readings, or the tritium purity) and also minimizes the time for setting the high voltage. Another scanning procedure tested during the FT campaign is the \textit{random-scan}, where the $qU_i$-values are set in random order. This scanning procedure is preferable to mitigate time-correlated effects, if present~\cite{Mertens:2012vs}.     

\begin{figure*}
\centering
\includegraphics[width=170mm]{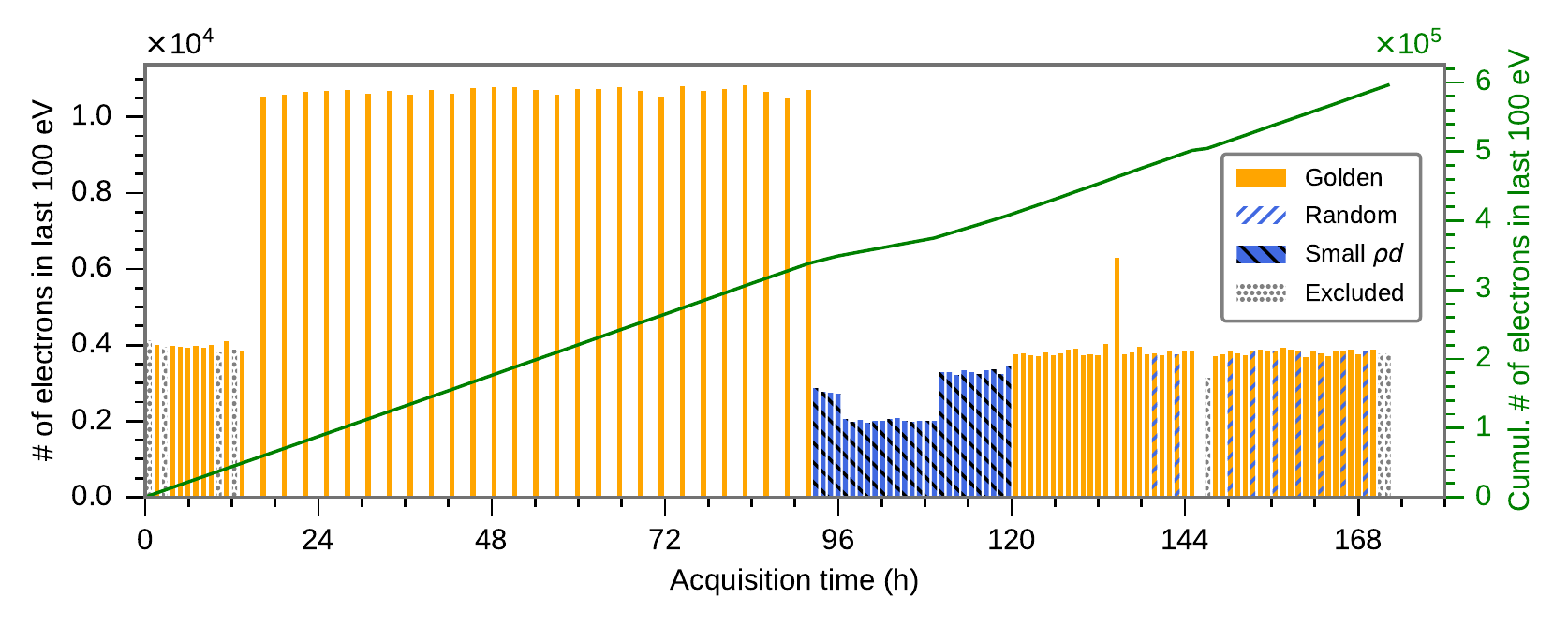}
\caption{Overview on performed scans during the FT campaign. The colored bars indicate the different measurement strategies. The height of the bar corresponds to the number of electrons recorded during a scan in an energy range of $E_0 - \SI{100}{\electronvolt} \leq E \leq E_0$. The higher bars correspond to \SI{3}{h} scans, while all others correspond to \SI{1}{h} scans (with the exception of one scan of about \SI{2}{h} at a time of $\sim\SI{130}{\hour}$). Gold: all (golden) scans used in the analysis (see \ref{sec:scan_selection}). Blue: special scans at different gas densities. Hatched-blue: scans, where the sub-scans were performed in random order. Grey: excluded scans. The green solid line indicates the cumulative number of electrons. In total, 168 hours (6 days) of scanning data were acquired, resulting in total statistics of about 0.6 million electrons. }
\label{fig:run_overview}
\end{figure*}

A measurement at a given retarding potential is called a \textit{sub-scan} and a full scan of all retarding potentials is defined as a \textit{scan}. The duration $t_{\text{scan}}=\sum_i t(qU_i)$ of a single scan was set to either one or three hours. The FT measurement entails 122 scans with a total measurement time for $\upbeta$-scans of 168 hours. Most of the scans were nominal \textit{up-} and \textit{down-scans} performed at \SI{100}{\percent} column density. A subset of scans was performed at \SI{20}{\percent}, \SI{50}{\percent}, and \SI{70}{\percent} column density to investigate the scattering of the $\upbeta$-electrons in the source. Another subset of scans was dedicated to test the technical feasibility of \textit{random scanning}. Figure~\ref{fig:run_overview} shows an overview of the acquired scanning data.

\section{Spectral analysis}
There are several challenges to the spectral analysis of the KATRIN data. 1) Due to various numerical integrals, the calculation of the integral $\upbeta$-electron spectrum is computationally intensive, which limits the flexibility with respect to the number of free parameters in the fit. 2) The analysis heavily relies on a precise description of the spectral shape including all relevant systematic effects and a robust treatment of systematic uncertainties. Any unaccounted-for effect and uncertainty can lead to systematic shifts of the deduced neutrino mass~\cite{Otten:2008zz}. 3) The KATRIN experiment acquires data in a sequence of $\mathcal{O}$(\SI{1}{h}) scans and the spectrum is recorded with $\mathcal{O}$(100) detector pixels. All these scans and pixels have to be combined in the final analysis without loss of information. In the following we describe the strategies on how to handle these challenges.

Two teams performed the analysis independently, each with its own spectrum calculation and analysis software.  The results presented in this work agreed within \SI{4}{\percent} percent of the total uncertainty, which gives a high confidence in our analysis tools.

\subsection{Calculation of the integral beta-decay spectrum}
\label{ssec:model}

The integral $\upbeta$-decay tritium spectrum is composed of two main parts: 1) the theoretical differential $\upbeta$-electron spectrum and 2) the experimental response function. The $\upbeta$-spectrum $R_\upbeta(E)$ is described by Fermi's theory
\begin{footnotesize}
\begin{equation}
 R_\upbeta(E) = C \cdot F(E,Z') \cdot p \cdot (E + m_{\text{e}}) \cdot (E_0 - E) \sqrt{(E_0-E)^2 - m^2_{\nu}},
\end{equation}
\end{footnotesize}%
where $C= \frac{G_F^2}{2\pi^3}\cos^2\Theta_C |M_{\text{nucl}}|^2$ with $G_{\mathrm{F}}$ denoting the Fermi constant, $\Theta_{\mathrm{C}}$ the Cabibbo angle, and $M_{\text{nucl}}$ the energy-independent nuclear matrix element. The $F(E, Z')$ represents the Fermi function with $Z'=2$ for the atomic number of helium, the daughter nucleus in this decay. $E$, $p$, and $m_{\text{e}}$ denote the kinetic energy, momentum, and mass of the $\upbeta$-electron, respectively. $E_0$ is the kinematic endpoint, i.e. the maximum energy the electron can obtain for the case of zero neutrino mass. 

$m^2_{\nu} = \sum_{i=1}^{3} | U_{ei} |^2 \, m_i^2$ is the effective electron antineutrino mass, defined as the incoherent sum of the neutrino mass eigenstates $m_i$, weighted by the squared absolute values of the respective elements in the Pontecorvo–Maki–Nakagawa–Sakata (PMNS) neutrino mixing matrix $U_{ei}$. $m^2_{\nu}$ is the observable of the KATRIN experiment.

After the $\upbeta$-decay of tritium in a DT molecule, the daughter molecule $\mathrm{^3HeD^+}$ can end up in an electronic ground state or excited state, each of which is broadened by rotational and vibrational excitations of the molecule~\cite{Bodine2015}. As a consequence, this excitation energy reduces the available kinetic energy for the electron, and thus, the differential $\upbeta$-electron spectrum is a superposition of spectra, corresponding to all possible final states. Each individual spectrum is weighted by the probability to decay into a certain final state and its spectral endpoint is reduced by the corresponding final-state energy.

The experimental response function 
\begin{footnotesize}
\begin{equation}
 f_\mathrm{calc}(E, qU_i) = \int_0^E T(E-\epsilon, qU_i) \left(P_0\,\delta(\epsilon) + P_1\,f(\epsilon) + \right.\\ 
 \left.P_2\,(f\otimes f)(\epsilon) + ...\right) \, \dd \epsilon,\label{eq:response}
\end{equation}
\end{footnotesize}%
is the probability of an electron with a starting energy $E$ to reach the detector. It combines the transmission function $T$ of the main spectrometer and the electron's energy losses $\epsilon$ in the source. The transmission function $T$ determines the resolution of the main spectrometer and is governed by the magnetic fields at the starting position of the electron, the maximum field in the beamline, and the magnetic field in the spectrometer's analyzing plane. Energy losses due to inelastic scattering with the tritium molecules in the source are described by the product of the $s$-fold scattering probabilities $P_s$ and the energy-loss function $f(\epsilon)$ convolved $(s-1)$ times with itself. In the case of no scatterings no energy is lost, which is expressed by the Dirac $\delta$-function $\delta(\epsilon)$. 

Synchrotron energy losses of $\upbeta$-electrons in the high magnetic field in the source and transport section are included as a correction to the transmission function. Furthermore, Doppler broadening due to the finite motion of the tritium molecules in the source is emulated as a broadening of the molecular final-state distribution. Finally, radiative corrections are included in the differential $\upbeta$-electron spectrum. The response function is slightly modified due to the dependence of the path length (and therefore effective column density) on the pitch angle of the $\upbeta$-electrons~\cite{Kleesiek:2018mel}. This effect is not taken into account in this analysis. The resulting effect on the measured endpoint, however, is small compared to the uncertainties of the electric potential of the source, as detailed in section~\ref{ssec:endpoint}. 

The spectrum calculation code, used in this work, is described in Refs.~\cite{thesis-lisa,thesis-christian} \footnote{Note that fit values may differ from those reported in this work since an early version of the data selection and systematics was employed at that time.}. A very detailed description of the full spectrum and instrument response calculation can be found in Ref.~\cite{Kleesiek:2018mel}.

The total rate $R_\mathrm{calc}(qU_i)$ at a given retarding energy $qU_i$ is given by  
\begin{footnotesize}
\begin{equation}
\label{Ntheo}
R_\mathrm{calc}(qU_i) = A_{\mathrm{s}} N_{\mathrm{T}} \int_{qU_i}^{E_0}R_\upbeta(E) f_\mathrm{calc}(E, qU_i) \ \dd E + R_{\mathrm{bg}},
\end{equation}
\end{footnotesize}%
where $N_{\mathrm{T}}$ is the signal normalization, which includes the number of tritium atoms in the source, the maximum acceptance angle and the detection efficiency. $A_{\mathrm{s}}$ is a free parameter in the fit and $R_{\mathrm{bg}}$ denotes the retarding-potential-independent background rate~\cite{Trost2019_1000090450}.

\subsection{Observed endpoint}
\label{ssec:endpoint}
The endpoint observed by the KATRIN experiment is influenced by the difference between the electric potential at the starting position of the $\upbeta$-electron $\Phi_{\text{WGTS}}$ and the work function $\Phi_{\text{MS}}$ of the main spectrometer, and is therefore not identical to the physical kinematic endpoint $E_0$. This observed endpoint
\begin{equation}
    E^{\text{fit}}_0 = E_0 + \Phi_{\text{WGTS}} - \Phi_{\text{MS}}
\end{equation}
is a free parameter in the spectral fit. The fitted endpoint $E^{\text{fit}}_0$ is related to the experimental $Q$-value for DT by taking into account the molecular recoil\footnote{A subtlety of this KATRIN analysis is that the final-state distributions for each tritium isotopologue are shifted to compensate for the mass-dependent recoil energies. Consequently, independently of which tritium isotopologue is present in the measurement, the fitted endpoint $E^{\text{fit}}_0$ corresponds to the one expected for T$_2$. Accordingly, in equation~(\ref{eq:Qvalue}) we need to use $E_{\text{rec}} = E^{\text{T}_2}_{\text{rec}}$.} $E_{\text{rec}}$:

\begin{equation}
\label{eq:Qvalue}
    Q^{\text{obs}}(\text{DT}) = E^{\text{fit}}_0 + E_{\text{rec}} - \left( \Phi_{\text{WGTS}} - \Phi_{\text{MS}} \right) ~.
\end{equation}

$\Phi_{\text{WGTS}}$ depends on plasma effects in the source and the work function of the rear wall $\Phi_{\mathrm{RW}}$. During the FT campaign, the beam tube was terminated with a stainless steel gate valve (as opposed to the gold-plated rear wall used in the neutrino mass measurement in 2019), for which the work function was not measured. 
As a consequence, the source potential $\Phi_{\text{WGTS}}$ is only known with an accuracy of about \SI{1}{\electronvolt} in the FT campaign.

The determination of the main spectrometer work function can be performed by measuring the electron transmission from a well-characterized electron-gun~\cite{Behrens2017} at an accuracy of several tens of meV~\cite{Behrens2016PhD}. However, this instrument was not available during the FT campaign. Therefore, the uncertainty of $\Phi_{\mathrm{MS}}$ is at least \SI{250}{\milli\electronvolt} \cite{Behrens2016PhD}. 

As a result, we assume that $\Phi_{\text{WGTS}} = \Phi_{\text{MS}} \pm \SI{1}{\electronvolt}$, despite the fact that both the gate valve and the main spectrometer are made of stainless steel. 

The determination of the $Q$-value also relies on an accurate high voltage (HV) calibration. Based on recent calibrations of the high-precision voltage divider \cite{Thummler:2009rz}, we estimate the uncertainty of the absolute voltage of the main spectrometer of about $\SI{94}{\milli\electronvolt}$~\cite{Arenz:2018ymp}, which is negligibly small compared to the uncertainty of the source's electric potential.

The calculated $Q$-value is based on high-precision Penning-trap measurements, which provide the atomic mass difference of $\mathrm{^3He}$ and T~\cite{Myers2019}. The most recent measurement yields $\Delta m = m\,(\mathrm{T})-m\,(^3\mathrm{He})=18592.01\pm\SI{0.07}{\electronvolt}$~\cite{Myers2015}. By taking into account the molecular dissociation and ionization energies, $E_\mathrm{D}$ and $E_{\mathrm{ion}}$, which can be derived from the ground-state energies of the molecules \cite{Saenz:2000dul} and the single and double ionization energies, \cite{NIST_ASD} one obtains a $Q$-value of~\cite{Otten:2008zz} 

\begin{align}
Q^{\text{calc}}(\text{DT}) &= \Delta m - E_\mathrm{D}(\mathrm{DT}) + E_\mathrm{D}(\mathrm{^3HeD^+}) - E_{\mathrm{ion}}(\mathrm{T}) \\&= 18575.71 \pm\SI{0.07}{\electronvolt}~. \label{eq:Qcalc}
\end{align}

\subsection{Fitting procedure}
In the standard KATRIN analysis, we consider four free parameters in the fit: the effective neutrino mass squared $m^2_{\nu}$, the signal normalization $A_{\mathrm{s}}$, background rate $R_{\mathrm{bg}}$, and the endpoint $E^{\text{fit}}_0$. As mentioned above, the accumulated statistics of the FT data are not sufficient to make a scientifically relevant statement about the neutrino mass. Instead, for the FT analysis the neutrino mass is fixed to zero and the endpoint $E^{\text{fit}}_0$ is treated as the parameter of interest.

In order to extract the physics parameters of interest, the model points $\vec{m}$, which may depend on several input parameters $\theta$, are fitted to the data points $\vec{d}$ by minimizing the negative Poisson Likelihood function
\begin{equation}
    -2\ln \mathcal{L} (\vec{d}|\theta) = 2\sum_i \left[m_i(\theta)-d_i+d_i \ln \left(\frac{d_i}{m_i(\theta)}\right)\right].
\end{equation}
For high-statistics spectra (for example, when many scans are combined) one can instead minimize the $\chi^2$ function:
\begin{equation}
\label{eq:chi2}
    \chi^2(\theta) = (\vec{d}-\vec{m}(\theta))\transpose C^{-1} (\vec{d}-\vec{m}(\theta)),
\end{equation}
where $C$ denotes the covariance matrix, describing the correlated and uncorrelated uncertainties of the model points $m_i$. Both statistical and systematic uncertainties can be embedded in the covariance matrix, see section~\ref{ssec:systematics-treatment}.

\subsection{Data combination}
\label{ssc:datacombination}
The FT data were used to test and optimize a diverse set of techniques for combining a large number of statistically independent spectra, recorded in different scans and with different detector pixels. As slow-control parameters may depend on time (for example, the source activity) and on the radial and azimuthal position in the beam tube (for example, the magnetic field), a subdivision of the data is necessary. As a first step of the analysis, the stability of fit parameters with respect to possible temporal and spatial variations is investigated. In the final analysis, however, a combined fit of all data is performed. Depending on the stability of slow-control parameters and on the required precision of the analysis, distinct options can be considered. 

\subsubsection{Scan combination}
To combine all scans we investigated the following possibilities:
\paragraph{Single-scan fit} In this method each scan is fitted individually. In this case, the spectrum calculation is initialized with the slow-control parameters of the corresponding scan. This procedure is important to observe the time dependence of fit parameters; however, it is not ideal for obtaining a final result based on all single-scan fits.
\paragraph{Stacking} Here, the counts in each sub-scan are added to construct a high-statistics single spectrum with the same number of data points $n_{\text{data-points}}=n_{\text{sub-scans}}$ as a single scan. As this method does not take into account scan-to-scan variations of slow-control parameters, a good time stability is required. Moreover, the stacking technique relies on a high reproducibility of the individual $qU_i$ settings. For the FT analysis, the effect of the underlying approximations of this method is negligible.
\paragraph{Appending} In order to avoid the requirement of reproducible $qU_i$ values, the data points of all scans can be combined in a single spectrum by simply appending them. In this case the single spectrum has $n_{\text{data-points}}=n_{\text{scans}} \cdot n_{\text{sub-scans}}$ data points. Again, in this technique, no scan-to-scan variation of slow-control parameters is taken into account in the spectrum calculation, and hence a high stability is required. 
\paragraph{Multi-scan fit} For exploiting the full potential of the KATRIN apparatus, scan-dependent (and potentially even sub-scan-dependent) information for all slow control and HV values are taken into account in the fit. In this way the requirements with respect to both HV reproducibility and scan-to-scan stability are significantly relaxed. However, the complexity of the spectrum calculation is significantly increased, and therefore this method has not been applied to the FT data.

\subsubsection{Pixel combination}
In the given configuration for the First Tritium campaign, the electric potential and magnetic field in the \SI{24}{\square\metre}-analyzing plane of the KATRIN main spectrometer are not perfectly homogeneous, but vary radially by about \SI{118}{\milli\volt} and \SI{1.75}{\micro\tesla}, respectively, and to a much smaller extent azimuthally. 

In order to account for this spatial dependence, KATRIN operates a 148-pixel detector (see layout in fig. \ref{Fig:pixel-map}). Each pixel has a specific transmission function and records a statistically independent tritium $\upbeta$-electron spectrum. In order to combine these spectra in the final analysis we can consider analogous options as for the scan combination:

\paragraph{Single-pixel fit} Each pixel is fitted individually. This procedure is important to observe the spatial dependence of fit parameters. However, obtaining a single final result by averaging the results of all pixels is not the preferred option, as the statistics of a single pixel is rather low and hence the fit values fluctuate severely. 

\paragraph{Uniform fit} The detector pixels are combined into a single pixel by adding all counts and assuming an average transmission function for the entire detector. This method is convenient and sufficient for several analyses, but the averaging of fields leads to a broadening of the spectrum and hence effectively worsens the energy resolution. 

\paragraph{Multi-pixel fit} For exploiting the full potential of the KATRIN apparatus, the multi-pixel fit can be applied, where all pixel-dependent spectra are fitted simultaneously. The fit assumes a common neutrino mass and endpoint but allows for pixel-dependent nuisance parameters, such as background, normalization, and HV-offsets. As a consequence, the number of free parameters is large: $n_\mathrm{free}=2+n_\mathrm{pixel}\cdot n_\mathrm{nuisance}\approx 446$ and hence the method is computationally expensive. A single fit with this number of free parameters takes on the order of 1 hour on a single CPU.

\subsection{Data selection}
\label{ssec:dataselection}
Data selection and combination are closely related. Specific ways of combining data impose certain stability and reproducibility requirements on the slow-control parameters. Depending on the analysis, we select a subset of all scans, a subset of detector pixels, and a certain fit range.

\paragraph{Scan selection} \label{sec:scan_selection}
Out of 116 scans, displayed in figure~\ref{fig:run_overview}, we excluded 34 scans for the following reasons: 1) 27 scans were performed at a different column density for testing purposes and are analyzed separately, 2) we exclude four scans where different HV setpoints were used than shown in figure~\ref{Fig:MTD}, 3) we exclude the last two scans and the first scan, as the DT concentration dropped by several percent. We define the resulting sub-set of 82 scans as the ``golden'' data set. 

For this golden data set the stacking technique leads to negligible errors on the endpoint $E^{\text{fit}}_0$. In order to test this, we simulate statistically-unfluctuated spectra, taking into account the scan-dependent slow-control parameters and the measured high-voltage values. We then fit this simulated data set, by stacking all scans and assuming average slow-control and high-voltage values. As a result, we find a negligible shift of \SI{10}{\milli\electronvolt} for the fitted endpoint $E^f{\text{fit}}_0$ compared to the Monte Carlo (MC) truth. This corresponds to \SI{4}{\percent} of the total 1-$\sigma$ uncertainty.

\paragraph{Pixel selection}
Out of the 148 pixels, the outer two detector rings (24 pixels) and three pixels of the third and forth outermost detector ring are not included in the analysis (see layout in fig. \ref{Fig:pixel-map}). Due to the alignment of the magnetic flux tube with the detector wafer and shadowing of the forward beam monitor, these pixels do not detect the full flux of $\upbeta$-electrons.

\begin{figure}[]
\centering
		\includegraphics[width=84mm]{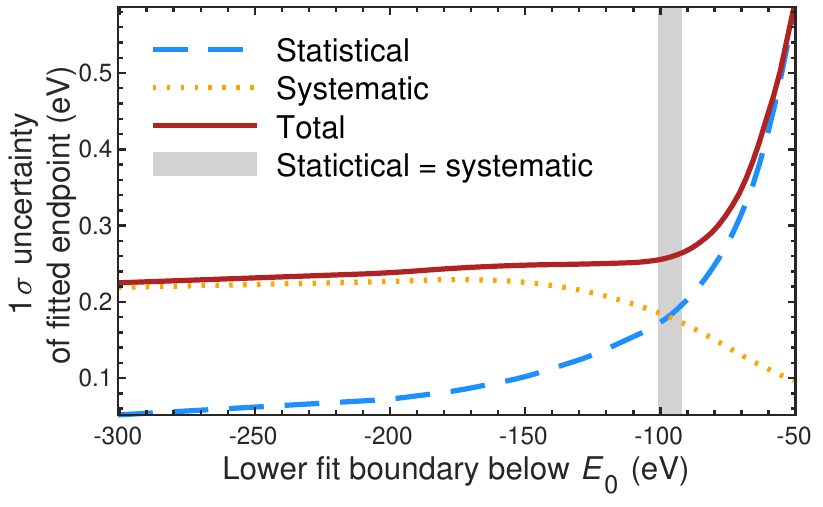}
		\label{Fig:statsys}
        \caption{Evolution of statistical and systematic uncertainties as a function of fit range. At a retarding energy of $qU=E_0 - \SI{100}{\electronvolt}$ a balance between systematic and statistical uncertainty is reached. We choose this range as the nominal energy range
        for this analysis.}
\end{figure}

\paragraph{Fit range selection}
The spectra were recorded over a large range down to \SI{1.6}{\kilo\electronvolt} below the endpoint. Depending on the specific analysis, a different range (i.e. set of sub-scans) can be included in the fit. Several systematic uncertainties increase further away from the endpoint, while the statistical uncertainty decreases. For the ``golden'' data set we choose a standard fit range with a lower limit of  $qU^{\min}=E_0 - \SI{100}{\electronvolt}$, since for this range the statistical and systematic uncertainties of the endpoint are of the same magnitude, see figure~\ref{Fig:statsys}.

\subsection{Systematic uncertainties}
Several calibration tools and measurements, such as a determination of the energy-loss function with a dedicated electron gun~\cite{Behrens2017} and a characterization of the plasma properties of the WGTS with a gaseous $^{83\mathrm{m}}$Kr source~\cite{V_nos_2014}, were not available at the time of the FT campaign. Moreover, the FT measurement interval extended much further into the spectrum (compared to a typical neutrino mass measurement), where several systematic uncertainties are enhanced. Consequently, the systematic uncertainties during the FT campaign do not fully reflect the final KATRIN systematic budget.

Nevertheless, the FT campaign allowed for a validation of our spectrum calculations and for testing of a set of methods to include systematic uncertainties for the subsequent neutrino mass analysis. In the following, the individual systematics and different ways of treating them in the analysis are discussed in detail.

\begin{table*}
	\caption{Budget of statistical and systematic uncertainties on the endpoint $E^{\text{fit}}_0$. The numerical values are based on the golden scan selection and the nominal fit range, as described in section~\ref{ssec:dataselection}. For this analysis the \textit{stacked-uniform} fit, as described in section~\ref{ssc:datacombination}, was applied. The column labeled ``uncertainty'' lists the $1\,\sigma$ uncertainties of the relevant input parameters. The column labeled ``impact on endpoint'' indicates the individual $1\,\sigma$ uncertainty contribution to the $E^{\text{fit}}_0$. In order to obtain the total uncertainty, all systematic effects were considered simultaneously, rather than adding the individual contributions in quadrature. For this analysis the systematics were included with the covariance matrix approach (see section~\ref{ssec:systematics-treatment}). For systematics labeled with ``on/off'', the maximum error estimation (see section~\ref{ssec:systematics-treatment}) was applied. It showed that the effect of a longitudinal gas  density profile, the effect of multiplicative theoretical corrections, as described in~\cite{Mer:2015a}, as well as the effect of analyzing the data with a \textit{stacked-uniform} fit have a negligible effect on the $E^{\text{fit}}_0$.}
	\label{tab:systematics}
	\centering
	\begin{tabular}{p{4cm}p{5cm}p{2.5cm}p{3cm}}
	\hline 
	                         Effect &  Description & $1\,\sigma$ uncertainty  & $1\,\sigma$ uncertainty of fitted endpoint (eV)  \\
	                   
    \hline
    \hline 
	\multirow[t]{2}{*}{Source scattering} & Column density    &   \SI{3}{\percent}          &    \SI{0.13}{}   \\
	                                      & Inel. scat. cross-section      &   \SI{2}{\percent}  &      \\
    \hline
    \multirow[t]{4}{*}{DT concentration fluctuation}        &  For single sub-scan (\SI{60}{\second})  & \SI{1.5}{\percent} &   \\
                                            &  For all scans combined (\SI{40000}{\second})   & \SI{0.08}{\percent} & \SI{0.03}{} \\	
    \hline
    \multirow[t]{5}{*}{Energy-loss function}        &  Excitation peak position $P_1$  & \SI{0.017}{\electronvolt}       & \SI{0.11}{} \\
	                                             &  Ionization peak position $P_2$  & \SI{0.18}{\electronvolt}        &\\
	                                     &  Excitation peak width $W_1$     & \SI{0.05}{\electronvolt}        &\\
	                                             &  Ionization peak width $W_2$     & \SI{0.13}{\electronvolt}        &\\	
	                                             &  Normalization $A$               & \SI{0.15}{\per\electronvolt} & \\  
	\hline
	\multirow[t]{3}{*}{Final-state distribution} &  Normalization                   & \SI{1}{\percent}  & \SI{0.08}{} \\
	                                             &  Ground-state variance           & \SI{1}{\percent}   &  \\
	                                             &  Excited-states variance         & \SI{3}{\percent}   &  \\
	\hline
	\multirow[t]{3}{*}{Magnetic fields}     &  Source            & \SI{2.5}{\percent} &   \SI{0.03}{}  \\
                          	                &  Analyzing plane   & \SI{1}{\percent}  &    \\
                                            &  Maximum field at pinch             & \SI{0.2}{\percent}  &     \\
	\hline
	Detector efficiency      &  Retarding potential dependence            &   \SI{0.1}{\percent}          &   \SI{0.03}{}         \\
	\hline
	Background          &  slope              & \SI{5}{\milli cps\per\kilo\electronvolt}     &    \SI{0.02}{}                  \\
	\hline
	Gas density profile      &  on/off             &                    &   $<\SI{0.01}{}$        \\
	Theoretical correction      &  on/off             &                    &   $<\SI{0.01}{}$         \\
	Stacking      &  on/off             &                    &   $<\SI{0.01}{}$         \\
	\hline
	\textbf{Total systematic uncertainty}  &   & &   \SI{0.19}{}       \\
	\textbf{Statistical uncertainty}  &              &                  &    \SI{0.17}{}                 \\
	\hline
	\textbf{Total uncertainty (stat. and syst.)}  &  &  & \SI{0.25}{}    \\
	\hline 
	\end{tabular}
	
\end{table*}

\begin{figure}[]
\centering
		\includegraphics[width=84mm]{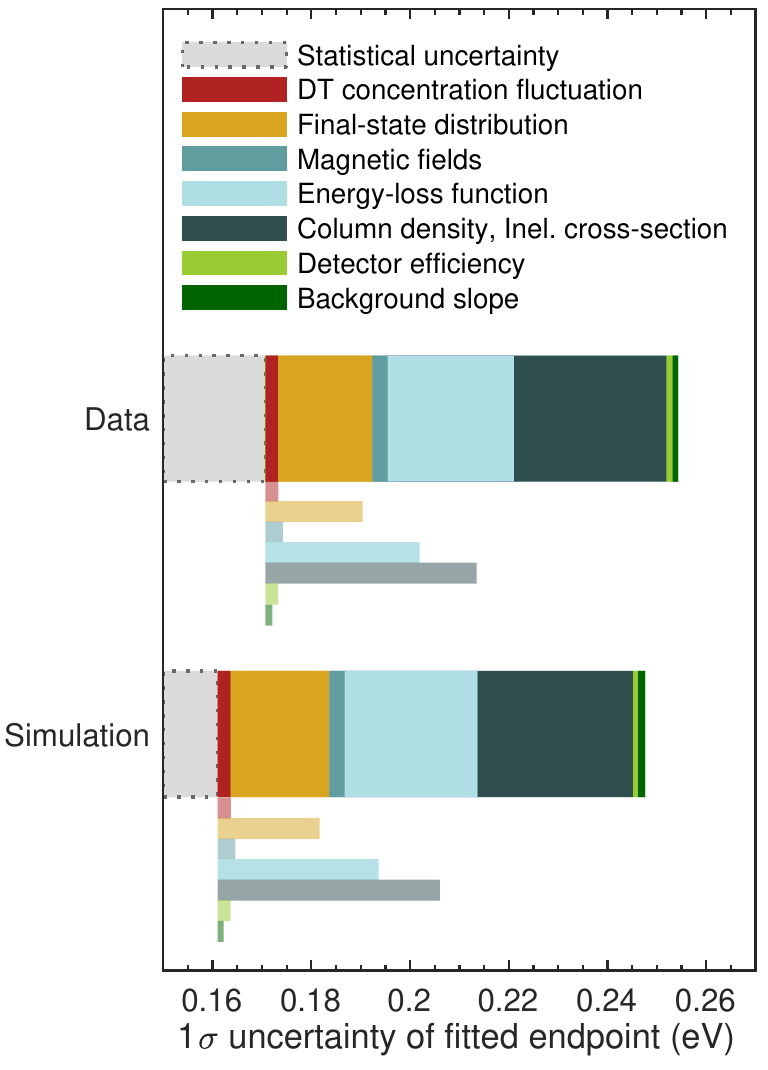}
		\label{Fig:systematics}
        \caption{Visual display of the systematic uncertainty breakdown as given in table~\ref{tab:systematics}. The analysis is based on the golden scan list and the nominal fit range, as defined in section~\ref{ssec:dataselection}. The data was analyzed with a stacked-uniform fit, as defined in section~\ref{ssc:datacombination}. Systematic uncertainties are included with the covariance matrix method.
        The upper set of bars shows the $1\,\sigma$ endpoint uncertainty based on the true data. The lower set of bars illustrates the expected $1\,\sigma$ uncertainty on the endpoint inferred from MC simulated data. A very good agreement is found. The individual bars (in light color) demonstrate the effect of each systematic uncertainty individually, as given in table~\ref{tab:systematics}. The stacked-bar (in darker color) displays the collective effect of all systematics when including them one-by-one in the fit. Note that due to correlations of uncertainties, the total uncertainty is not exactly given by the sum of the squared individual uncertainties.  }
\end{figure}

\subsubsection{Systematics budget}
Systematic uncertainties in KATRIN generally arise from uncertainties and instabilities of parameters, which enter into the calculation of the integral spectrum. Table~\ref{tab:systematics} summarizes the systematic uncertainty budget for the FT measurement; figure~\ref{Fig:systematics} graphically displays the impact of the individual systematic effects on the endpoint $E^{\text{fit}}_0$. In the following, the individual systematics will be described in detail.

\paragraph{Column density}
A major systematic effect for the FT measurement arises from the uncertainty of the column density. The column density $\rho d$ firstly determines the number of tritium atoms $N^{\text{tot}}$ in the source
\begin{equation}
\label{eq:NumberTritium}
    N^{\text{tot}} = \epsilon_{T} \cdot \rho d \cdot A,
\end{equation}
where $\epsilon_{T}$ is the tritium purity and $A$ is the cross sectional area of the WGTS. Secondly, the column density determines the scattering probability $P_s$ (see equation~(\ref{eq:response}) ) of electrons in the source~\cite{Kleesiek:2018mel}. In good approximation, the column density can assumed to be constant in radius~\cite{Kleesiek:2018mel}. 

Of relevance for the KATRIN analysis are 1) unaccounted-for variations of the total number of tritium atoms $N^{\text{tot}}$ during a scan and 2) the precise knowledge of the scattering probabilities $P_s$, and therefore the product of $\rho d \cdot \sigma_{\mathrm{inel}}$, where $\sigma_{\mathrm{inel}}$ is the cross-section for inelastic scattering of electrons off molecular deuterium (dominant isotopologue during the FT campaign). The precise absolute value of $N^{\text{tot}}$ is of minor relevance as it only influences the spectrum normalization and not its shape.

For the FT campaign, the stability of the column density was monitored via the gas flow into the WGTS, the buffer vessel pressure and the beam tube temperature. All three showed extremely small relative variations on the order of $10^{-5}$ on the time scale of minutes (sub-scan length). This variation is much smaller than the statistical uncertainty on the number of detected $\upbeta$-electrons, and therefore negligible.

The absolute column density was determined via the buffer vessel pressure combined with dedicated gas simulations \cite{Kuckert2018}. The corresponding systematic uncertainty is estimated to be $\sigma_{\rho d} = \SI{3}{\percent}$. For the cross-section $\sigma_{\mathrm{inel}}= \SI{3.65e-18}{\per\centi\metre\squared}$ of \SI{18.6}{\kilo\electronvolt} electrons on deuterium (based on~\cite{Liu1987}), we assume a conservative uncertainty of \SI{2}{\percent}. Finally, the product of column density and cross section depicts the dominant systematic uncertainty $\sigma_{\rho d \cdot \sigma_{\mathrm{inel}}} = \SI{3.6}{\percent}$ for the FT campaign.

For the neutrino mass measurements, KATRIN will use a dedicated electron gun~\cite{Behrens2017} to determine the scattering probabilities $P_s$ directly. An uncertainty of $\sigma_{\rho d \cdot \sigma_{\mathrm{inel}}} = \SI{0.1}{\percent}$ is targeted. 

\paragraph{Tritium concentration}
Together with the column density, the tritium concentration $\epsilon_{T}$ determines the total number of tritium atoms in the source, see equation~(\ref{eq:NumberTritium}). Here again, unaccounted-for variations of the tritium concentration are relevant as they can introduce distortions of the shape of the tritium spectrum.

During the FT measurements, the tritium concentration was constantly monitored by a Laser Raman system integrated into the inner loop system of the WGTS~\cite{Fischer:2012lfa}.

At the time of the FT campaign, the source gas molecules comprised only \SI{0.5}{\percent} tritium atoms, predominantly in the form of DT, therefore the relative statistical uncertainty of the Laser-Raman spectroscopic measurement was on the order of a few percent on time scales of minutes (sub-scan length). In the final fit, however, where all scans are combined, the statistical uncertainty on the DT concentration is reduced to $\sigma_{c(\mathrm{DT})}=\SI{0.08}{\percent}$.

In the design operation of KATRIN, the tritium purity of the source gas will be higher than \SI{95}{\percent}. In this case, the statistical uncertainty of the tritium purity measurement by the Laser Raman system will be significantly improved. The most relevant effect will then be the relative concentrations of the most abundant active gas isotopologues T$_2$, HT, and DT. As these different isotopologues have slightly different kinematic endpoints, their relative concentrations have an influence on the spectral shape in the energy range of interest for the neutrino mass. 

\paragraph{Energy-loss function}
The energy-loss function describes the probability of a \SI{18.6}{\kilo\electronvolt} $\upbeta$-electron to lose a certain amount of energy in a single inelastic scattering. For the analysis of the FT data the energy-loss function measured by the \textit{Troitsk nu-mass} experiment~\cite{Abdurashitov:2016nrv} with H$_2$ and D$_2$ is used. The function is described by an empirical model containing six parameters, namely the position $P$ and width $W$ of the excitation (index 1) and ionization (index 2) peaks as well as the normalizations $N$ and $A$.

\begin{equation} 
\label{eq:ELoss}
f(\varepsilon)= N \cdot \begin{cases} A \cdot \exp\left(-\frac{2(\varepsilon-P_1)^2}{W_1^2}\right) & \textrm{for } \varepsilon \leq \varepsilon_c \\
 \frac{W_2^2}{W_2^2+4(\varepsilon-P_2)^2} & \textrm{for }  \varepsilon > \varepsilon_c \\
\end{cases}
\end{equation}

We use the parametrization and correlated uncertainties as quoted in~\cite{Abdurashitov:2016nrv} averaged over both isotopologues, as can be seen in table~\ref{tab:systematics}.

For subsequent neutrino mass measurements, the energy-loss function will be precisely determined by the KATRIN experiment itself by means of a pulsed electron gun and operating the experiment in the time-of-flight mode~\cite{Bonn1999, Behrens2017}. A publication on the first successful measurements of the energy-loss function with the KATRIN apparatus is currently in preparation.

\paragraph{Magnetic fields}
The entire KATRIN beamline is composed of about sixty super-conducting and normal-conducting magnets. The source magnetic field $B_{\text{source}}$, the maximum magnetic field $B_{\text{max}}$, and the magnetic field in the analyzing plane $B_{\text{ana}}$ determine the shape of the transmission function, the maximum angular acceptance, and the energy resolution of the main spectrometer. With a magnetic field setting of $B_{\text{source}}=\SI{2.52}{\tesla}$, $B_{\text{max}}=\SI{4.2}{\tesla}$, and $B_{\text{ana}}=\SI{6.3d-4}{\tesla}$, an energy resolution of $\Delta E = \SI{18575}{\electronvolt} \cdot \frac{B_{\text{ana}}}{B_{\text{max}}}= \SI{2.8}{\electronvolt}$ was achieved during the FT campaign. 

We assume uncertainties of the magnetic fields of $\sigma_{\text{B}_{\text{source}}} = \SI{2.5}{\percent}$, $\sigma_{\text{B}_{\text{ana}}} = \SI{1}{\percent}$ and $\sigma_{\text{B}_{\text{max}}} = \SI{0.2}{\percent}$. These values are estimated based on comparisons of simulations with the KATRIN software Kassiopeia~\cite{Furse:2016fch} and measurements with Hall sensors and precision magnetic field sensors~\cite{Erhard:2017htg, Heizmann2019_1000093536}. 

The strongest magnet in the KATRIN beamline, the pinch magnet which defines ${\text{B}_{\text{max}}}$, is running in persistent mode and is therefore extremely stable at about 40 ppm over a period of 60 days. The stability of the other magnets, defining $\text{B}_{\text{source}}$ and $\text{B}_{\text{ana}}$, is monitored with precise magnetometers and electric current sensors, respectively. During the FT campaign a stability at the \SI{0.1}{\percent} level is observed. This stability meets the requirements of the final KATRIN design and contributes a negligible systematic effect for the FT analysis.
A detailed description of the monitoring of the magnet system of KATRIN and its performance can be found in~\cite{Arenz:2018jpa}.

Future dedicated measurements with an electron gun are expected to improve the accuracy of the source magnetic field by one order of magnitude. Furthermore, the application of a complex magnetic field sensor system~\cite{Letnev:2018fkq} will prospectively improve the uncertainty of the analyzing plane magnetic field by a factor of five.

\paragraph{Electric Potentials}
Uncertainties of the absolute value of the electric potentials in the source and spectrometer are absorbed by the fitted endpoint $E^{\text{fit}}_0$, as described in detail in section~\ref{ssec:endpoint}. These uncertainties do not affect the neutrino mass measurement; however, they do need to be taken into account when comparing $E^{\text{fit}}_0$ to the true kinematic endpoint and the Q-value of the spectrum.

More relevant for the spectral analysis are spatial and temporal fluctuations of electric potentials. A short-term ($<$ time of sub-scan) time fluctuation of the source and/or spectrometer potential leads to a broadening of the $\upbeta$-electron spectrum~\cite{Thummler:2009rz}. A longitudinal variation of the source electric potential analogously leads to a distortion of the observed $\upbeta$-electron spectrum~\cite{Kuckert:2018kao}. 

During the FT campaign, an excellent HV stability of $< \SI{40}{\milli\volt}$ during a sub-scan was observed, which is better than the requirements for the final neutrino mass measurement ($< \SI{60}{\milli\volt}$). Moreover, due to the dilute amounts of tritium gas, source plasma inhomogeneities are expected to be negligible. Consequently, the associated systematic uncertainties are assumed to be negligibly small for the FT campaign.

\paragraph{Final-State Distribution}
An unavoidable systematic effect stems from the fact that KATRIN uses molecular tritium (as opposed to atomic tritium). The rotational and vibrational excited states of the molecules inherently lead to a broadening of the $\upbeta$-electron spectrum. However, the more severe effect for KATRIN is a possible theoretical uncertainty on the description of the final-state distribution. 

At the time of the analysis there was no final-state distribution available for the most abundant tritium-containing isotopologue DT during FT campaign. Therefore, it was decided to adopt the final-state distribution of the HT isotopologue calculated by Saenz et al.~\cite{Saenz:2000dul}. The isotope effects, i. e.\ the influence of the broadening of the initial vibrational ground-state wavefunction and the recoil on the mean excitation energy and variance of the final-state distribution is discussed in
\cite{Saenz1997, Saenz:2000dul, Bodine2015}. With the conservative assumption of \SI{1}{\percent} uncertainty on the relative normalization between ground and excited states, \SI{1}{\percent} uncertainty on the variance of the ground-state distribution, and \SI{3}{\percent} uncertainty on the excited-state distribution the adopted final-state distribution for HT (instead of DT) is still found to be sufficiently accurate for the present purpose. The analysis of future runs of KATRIN requires the calculation of a more appropriate and accurate final-state distribution. Such calculations are currently in progress.

\paragraph{Detector Efficiency}
Since the KATRIN focal plane detector counts electrons as a function of the retarding potential, its retarding-potential-dependent detection efficiency is of major importance. The absolute efficiency, on the other hand, impacts only the total statistics, but does not alter the shape of the spectrum.

The following effects can lead to a retarding-potential-dependent detector efficiency: 

a) The focal plane detector provides a moderate energy resolution of about \SI{3}{\kilo\electronvolt} (full-width-half-maximum). Consequently, the electrons are counted in a wide region of interest (ROI) of $\SI{14}{\kilo\electronvolt} \leq E + qU_{\text{PAE}} \leq \SI{32}{\kilo\electronvolt}$, where $E$ is the $\upbeta$-electron energy and $U_{\text{PAE}}=\SI{10}{\kilo\electronvolt}$ is the post-acceleration voltage applied to the detector. The recorded differential energy spectrum changes slightly as the retarding potential changes. For a fixed ROI, this leads to a slight over/under counting of events. At $qU = E_0 - \SI{100}{\electronvolt}$ this effect amounts to $\delta_{\text{ROI}} = 1 - \epsilon_{\text{ROI}} = \SI{0.2}{\percent}$ with $\sigma_{\delta_{\text{ROI}}}=\SI{0.16}{\percent}$.

b) The rate at the detector varies with the retarding potential, and so does the probability of pile-up (pu). After pile-up correction, this effect alters the detection efficiency at $qU = E_0 - \SI{1}{\kilo\electronvolt}$ by $\delta_{\text{pu}} = 1 - \epsilon_{\text{pu}} = \SI{0.02}{\percent}$, with $\sigma_{\delta_{\text{pu}}}=\SI{18}{\percent}$.

c) Electrons backscattered (bs) from the detector surface can be lost if they overcome the retarding potential of the main spectrometer a second time. Consequently, as the retarding potential is lowered, the probability of lost electrons increases. At \SI{1}{\kilo\electronvolt} below the endpoint, this leads to a change of the detector efficiency of $\delta_{\text{bs}} = 1 - \epsilon_{\text{bs}} = \SI{0.15}{\percent}$, with $\sigma_{\delta_{\text{bs}}}=\SI{50}{\percent}$.

For the FT measurement, a pixel-dependent region-of-interest ($\epsilon_{\text{ROI}}$) and pile-up ($\epsilon_{\text{pu}}$) correction was taken into account. The corrections at the nominal range of $qU_i \geq E_0 - \SI{100}{\electronvolt}$ are significantly smaller than at $qU_i \geq E_0 - \SI{1}{\kilo\electronvolt}$. As a conservative approach, we consider a sub-scan to sub-scan independent uncertainty of the detector efficiency of \SI{0.1}{\percent}. For the final neutrino mass analysis the effect will be even smaller, as the scanning range will be reduced to about $qU_i \geq E_0 - \SI{40}{\electronvolt}$.

\paragraph{Background}
During the FT measurement an average background rate of \SI{350}{\milli cps} was observed. An increasing background rate moves the neutrino mass signature away from the endpoint, where the signal is weaker and systematic effects become more dominant. Several means to reduce the background rate to $< \SI{100}{\milli cps}$ are currently under investigation.

A fraction of the background arises from Rn-219 and Rn-220 decays in the volume of the main spectrometer and subsequently magnetically stored electrons. Through ionization of residual gas, this primary stored electrons creates numerous low-energy secondary electrons, which can reach the detector and create background~\cite{Frankle:2011xy, Wandkowsky:2013una, Wandkowsky:2013vna, Mertens:2012vs, Goerhardt:2018wky}. These background events are correlated in time, and hence the total background rate is not Poisson distributed. The observed broadened rate distribution, which can be described by a Gaussian-broadened Poisson distribution, is of major importance for the sensitivity of the KATRIN experiment~\cite{Mertens:2012vs}.

Based on sub-scans above the endpoint during the FT campaign, a Gaussian broadening with a variance of $\sigma^2=4.3^{+5.5}_{-4.8} \cdot 10^{-5}$~cps$^{2}$ was found. Due to the large uncertainty, this result is compatible with no Gaussian broadening. If we consider $\sigma^2=\SI{4.3e-5}{\cps\squared}$ (corresponding to a broadening by \SI{3}{\percent}) the uncertainty on the fitted endpoint would be enlarged by \SI{0.02}{\electronvolt}, which would depict a minor contribution in the systematic budget. In future measurement campaigns more sub-scans above the endpoint are planned to determine the non-Poisson nature of the background with higher accuracy.

A second relevant property of the background is a possible retarding-potential dependence. Several long-term measurements did not reveal any indication of a slope and thus point at a limit of $< \SI{5.3}{\milli cps \per\kilo\electronvolt}$ at $1\,\sigma$. For the analysis of the FT spectra we treat the slope as constrained systematic uncertainty.

\subsubsection{Treatment of systematics}
\label{ssec:systematics-treatment}
A main objective of the FT campaign was to explore suitable techniques to include systematic uncertainties. The following techniques were successfully applied: nuisance parameter method, covariance matrix method, Monte Carlo propagation of uncertainties, and a simple maximum error estimation. In this paper we discuss each technique in a concise fashion. A more detailed discussion of the methods will follow in a separate publication.

\paragraph{Nuisance parameters}
An elegant method to treat uncertainties of systematic parameters is to include them as additional free parameters in the fit, with the option of constraining their value with a nuisance term in the likelihood function to a range provided by external information. 

This method is applied in the KATRIN data analysis at least for the signal normalization $A_{\mathrm{s}}$, the background normalization $R_{\mathrm{bg}}$, and the endpoint $E^{\text{fit}}_0$. Other systematic parameters can also be treated as nuisance parameters. This technique was applied for example for the column density and background slope.However, if the number of free parameters is too large, the minimization of the likelihood function can become extremely computationally challenging. 

\paragraph{Covariance matrix}
Another less computationally intensive way to include uncertainties of input parameters is via the so-called multi-sim covariance matrix method~\cite{DAgostini:1993arp, Barlow:213033, thesis-lisa}. Here, the spectrum prediction is computed thousands of times while varying the systematic parameters according to a given distribution each time. In this way, the variance and also the covariance of the spectral data points, caused by the uncertainty of the systematic parameter, is extracted. The full covariance matrix, $C$, is then included in the $\chi^2$-function as can be seen in equation~(\ref{eq:chi2}).

This approach is particularly applicable for large counting statistics, in which case the application of the $\chi^2$ minimization is justified. This, in turn, requires stacking spectra of different scans or pixels in order to accumulate sufficient statistics per retarding potential. 

\paragraph{Monte Carlo propagation}
A promising method is based on Monte Carlo propagation of uncertainties~\cite{Cousins:1991qz, Cowan:2010js, Harris_2014, thesis-christian}. Here, the full fit is executed thousands of times while varying the systematic input parameters according to a given distribution in each fit. The widths of the resulting distributions of the fit parameters provide a measure of the systematic uncertainty of this fit parameter. To extract the maximum information from the data, each fit result is weighted with the likelihood to obtain the measured data points, given the particular choice of systematic parameter. In order to simultaneously treat statistical and all systematic uncertainties, each fit is performed on a statistically fluctuated MC-copy of the true data set, where fluctuations can entail Poisson rate fluctuations, non-Poissonian background fluctuations, correlated tritium activity fluctuations, and HV-variations from sub-scan to sub-scan. 

This method does not require large statistics and avoids the technical difficulties that would arise when treating all uncertainties with free nuisance parameters.

\paragraph{Maximum error estimation}
The maximum error estimation, or \textit{shift-method}, is a simple approach to access the impact of a neglected effect in the spectrum calculation. Here, a Monte Carlo data set is generated based on a spectrum model \textit{A}, which is then fitted with another spectrum model \textit{B}, where a certain effect is neglected. The resulting shift of the fitted parameter of interest (here the endpoint $E^{\text{fit}}_0$) with respect to the Monte Carlo truth, indicates whether or not the effect needs to be taken into account. 

This approach was used for the FT analysis to evaluate to which level of accuracy the KATRIN spectrum is required to be calculated. Using this method, it could be shown that neglecting effects such as a segmentation of the WGTS to take into account the longitudinal and radial gas profile is justified for the FT campaign.

\section{Results}
As the FT data provides no relevant statistical sensitivity to the neutrino mass, the endpoint $E^{\text{fit}}_0$ was treated as the parameter of interest in this analysis. The main focus of this measurement campaign was to use the endpoint value 1) to compare different analysis strategies, 2) to evaluate the independence of the fit result on the column density, scanning strategy, and fit range, and 3) to demonstrate time- and spatial-stability of the fits.

Combining all data, by stacking the golden scans, treating the golden pixels as a single effective pixel (uniform fit), and performing a fit in the nominal range of $qU_i > E_0 - \SI{100}{\electronvolt}$ we find an endpoint of 
    \begin{eqnarray}
E^{\text{fit}}_0(\mathrm{DT}) &=& 18574.39 \pm 0.17 (\text{stat}) \pm 0.19 (\text{sys})~\text{eV} \nonumber \\ 
&=& 18574.39 \pm 0.25 (\text{tot})~\text{eV},
\end{eqnarray} 
where the systematic uncertainty was obtained via the covariance matrix method. This corresponds to an endpoint for T$_2$ of $E^{\text{fit}}_0(\mathrm{T_2})=18574.73\pm 0.25 (\text{tot})~\text{eV}$ - taking into account shifts from recoil and differences in the electronic ground states between DT/T$_2$ and $^3$HeD$^+$/$^3$HeT$^+$.

Based on equation~(\ref{eq:Qvalue}) we can derive a $Q$-value for DT of $Q^{\text{obs}}(\text{DT})=18576.45\pm \SI{1}{\electronvolt}$, where the large uncertainty mainly stems from the uncertainty of the work function of the rear end of the beam tube during the FT campaign. The value is in agreement with the calculated $Q^{\mathrm{calc}}$-value of $Q(\text{DT})=18575.71\pm\SI{0.07}{\electronvolt}$ (see Eq. (\ref{eq:Qcalc})).

It is important to note that in upcoming measurement campaigns, a gold-plated rear-wall will be terminating the KATRIN beamline, which exhibits a significantly different work function compared to the stainless steel gate-valve used during the FT campaign. Moreover, a much higher tritium activity will be present in the source, which prospectively leads to the formation of a plasma potential. As a consequence, the source electric potential, and hence the measured endpoint $E^{\text{fit}}_0$ will prospectively differ significantly in future KATRIN measurements compared to the value reported here.

\begin{figure*}[]
    \subfigure[]{\includegraphics[width=84mm]{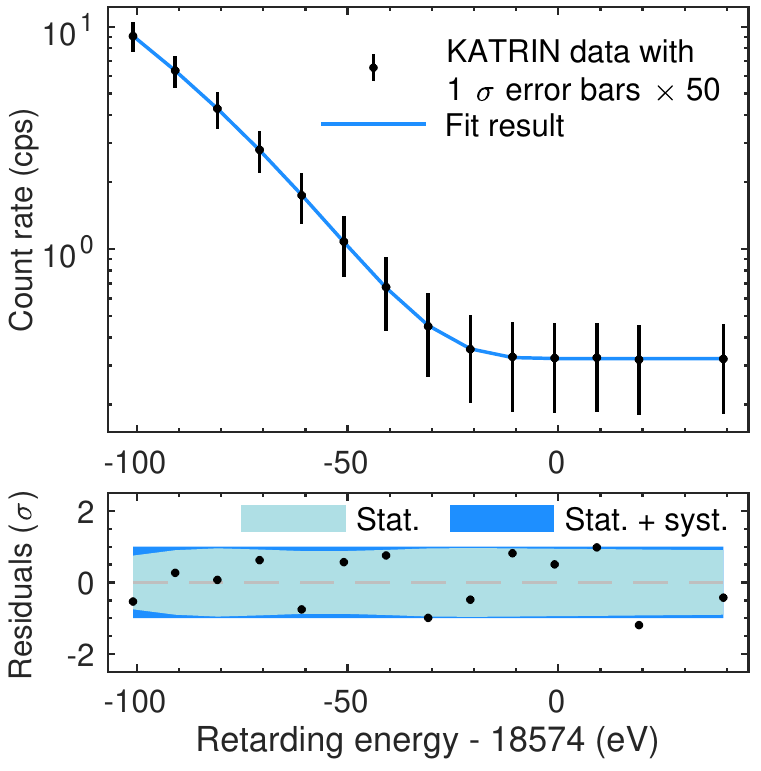}}\subfigure[]{\includegraphics[width=84mm]{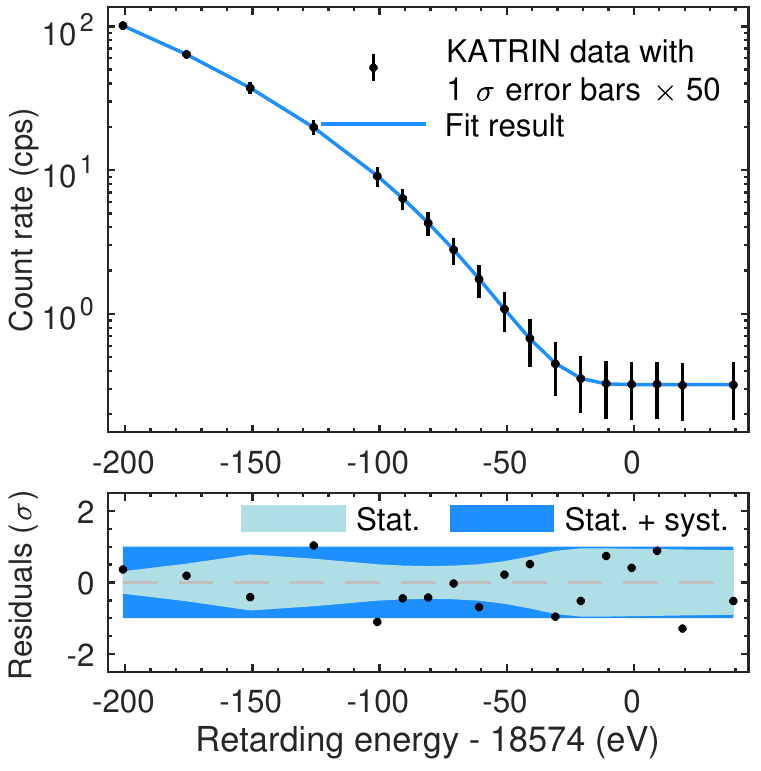}}
    \subfigure[]{\includegraphics[width=84mm]{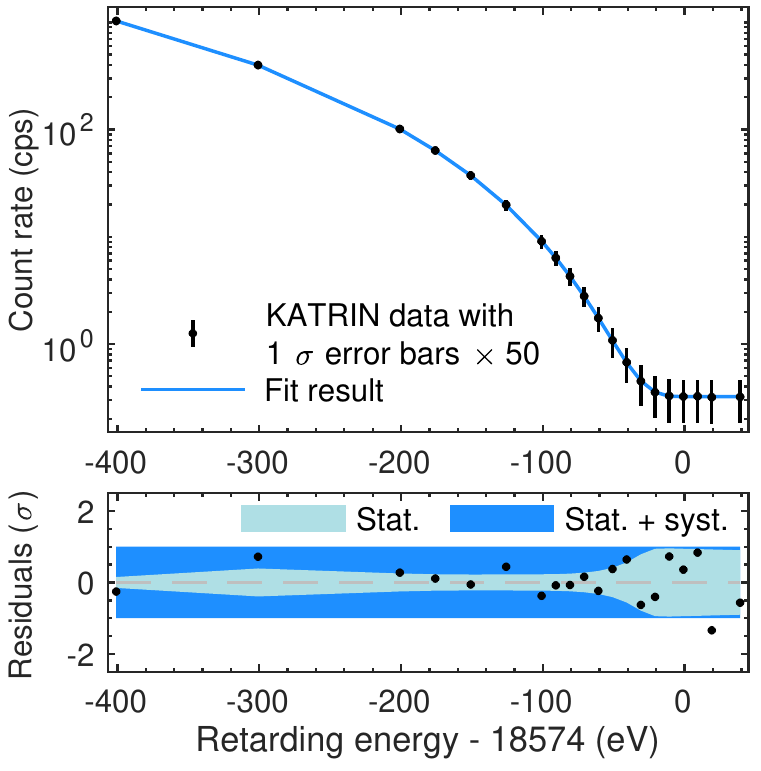}}
    \caption{Fit of the golden data selection in three selected fit ranges using the covariance matrix approach. The error bars are increased by a factor of 50 to make them visible. The residuals are normalized to the total uncertainty. The light-blue area indicates the statistical and the dark-blue area the systematic contribution to the total uncertainty. In this display of the systematic uncertainty band, only the diagonal entries of the covariance matrix are shown. a) Nominal fit range of $qU_i \geq E_0 - \SI{100}{\electronvolt}$, $\chi^2 = 7.9$ (11 dof). b) Mid-extended range to $qU_i \geq E_0 - \SI{200}{\electronvolt}$, $\chi^2 = 12.7$ (15 dof). c) Large-extended range to $qU_i \geq E_0 - \SI{400}{\electronvolt}$, $\chi^2 = 13.8$ (17 dof).}
    \label{fig:fits}
\end{figure*}

Figure~\ref{fig:fits} shows the fit result for three selected fit ranges down to $qU_i > E_0 - \SI{400}{\electronvolt}$. The excellent goodness of the fit in all cases indicates a good understanding of the spectral shape even far beyond the standard KATRIN energy range of $qU_i > E_0 - \SI{40}{\electronvolt}$.

\begin{figure}[]
\centering
		\includegraphics[width=84mm]{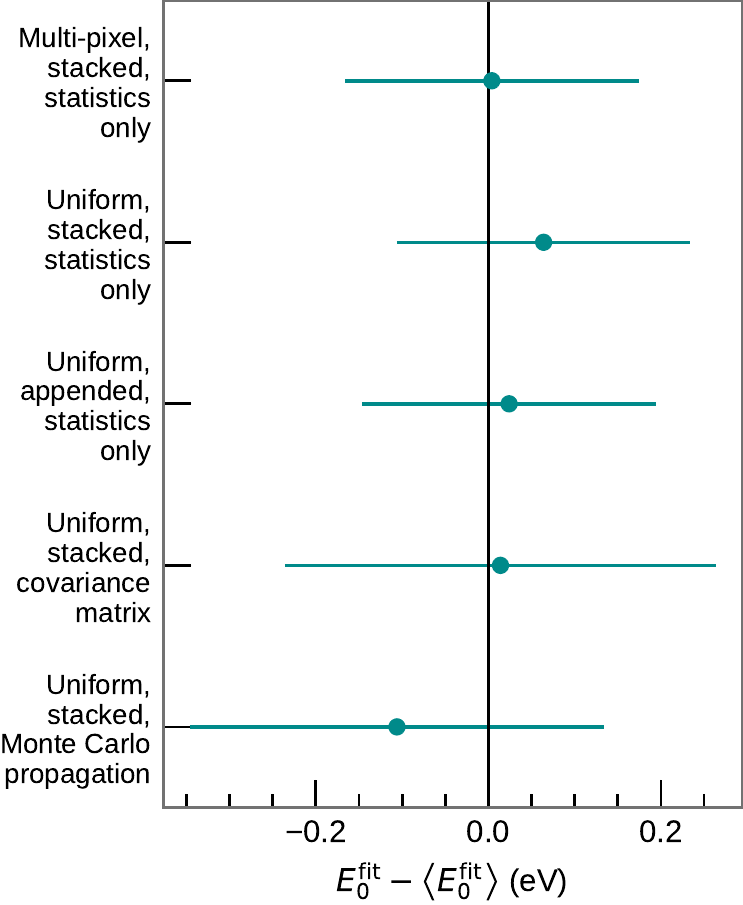}
		\label{Fig:methodcomparison}
        \caption{Fitted endpoint $E^{\text{fit}}_0$ for different analysis methods as described in section~\ref{ssc:datacombination}. The first and second data point compare two different ways of combining detector pixels (\textit{multi-pixel} and \textit{uniform} treatment of pixels). The second and third data points compare distinct ways of combining scans (\textit{stacking} and \textit{appending} of sub-scans). The lower two data points correspond to two different ways of including systematic uncertainties (\textit{covariance matrix method} and \textit{MC propagation} of uncertainties). The results obtained with different methods are in good agreement. It is expected that the best-fit value depends slightly on the way systematic uncertainties are treated.}
\end{figure}

As can be seen in figure~\ref{Fig:methodcomparison}, within the total uncertainty the results of the different analysis techniques show good agreement. On the one hand, this illustrates the high stability of the system, which makes it possible to apply simplifications in the analysis, such as the stacking of scans. On the other hand it shows the readiness of the more advanced techniques, such as a simultaneous fit of all pixels with a large number of free parameters.

\begin{figure*}[]
    \subfigure[]{\includegraphics[width=84mm]{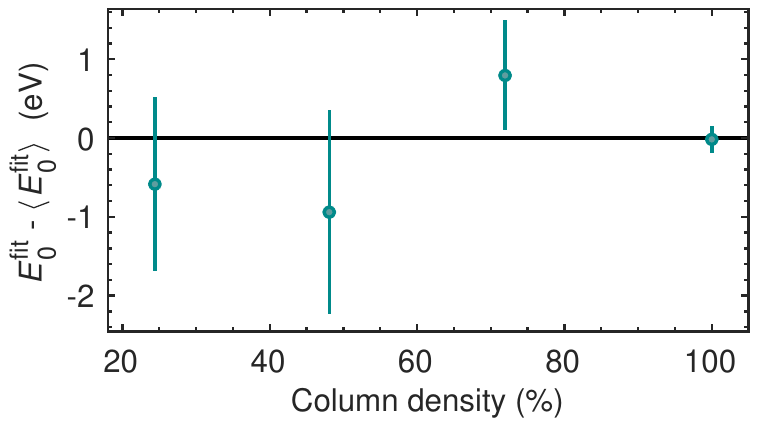}}\\
    \subfigure[]{\includegraphics[width=84mm]{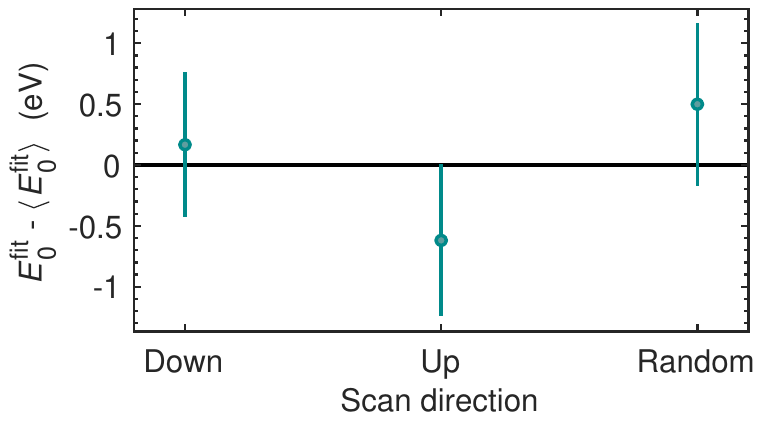}}\\
    \subfigure[]{\includegraphics[width=84mm]{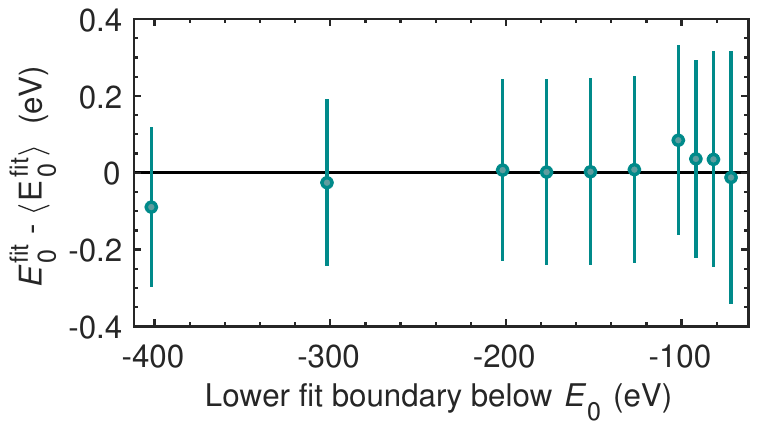}}
    \caption{Fitted endpoint $E^{\text{fit}}_0$ for different experimental conditions. Here, the data was analyzed with a stacked-uniform fit, as defined in section~\ref{ssc:datacombination} and systematic effects were included via the covariance matrix approach, as defined in section~\ref{ssec:systematics-treatment}. a) $E^{\text{fit}}_0$ for different column densities.  b) $E^{\text{fit}}_0$ for different scanning strategies. c) $E^{\text{fit}}_0$ as a function of fit range. Here the upper fit boundary is fixed to \SI{40}{\electronvolt} above the endpoint, and the lower fit boundary takes values between \SI{-400}{\electronvolt} and \SI{-60}{\electronvolt} below the endpoint. In each panel, the black line was calculated as weighted mean. }
    \label{fig:ConditionComparison}
\end{figure*}

Another important outcome of the campaign was the demonstration that the fitted endpoint $E^{\text{fit}}_0$ does not depend on the column density in the source, see figure~\ref{fig:ConditionComparison}a. For this purpose, dedicated scans at \SI{20}{\percent}, \SI{50}{\percent}, \SI{70}{\percent}, and \SI{100}{\percent} column density were performed. The independence of the fitted endpoint $E^{\text{fit}}_0$ on the column density gives confidence in a good understanding of the scattering processes in the source. 

Another set of dedicated scans was performed to check  whether the fit parameters depended on the scanning mode. Fitting the parameter $E^{\text{fit}}_0$ for a set of up-, down-, and random scans individually we find no dependence within the uncertainty, see figure~\ref{fig:ConditionComparison}b.

An important test of the correctness of our spectrum calculation is the $qU_i$-scan. Here, we check the parameter stability with respect to the fit range. Figure~\ref{fig:ConditionComparison}c shows that $E^{\text{fit}}_0$ has indeed no statistically significant dependence on the fit range between $qU_i \geq E_0 - \SI{400}{\electronvolt}$ and $qU_i \geq E_0 - \SI{60}{\electronvolt}$. As the individual fit results are not statistically independent from each other, a Monte Carlo study, was performed, which confirms the independence of the fit result on the fit range.

\begin{figure}
\centering
		\includegraphics[width=84mm]{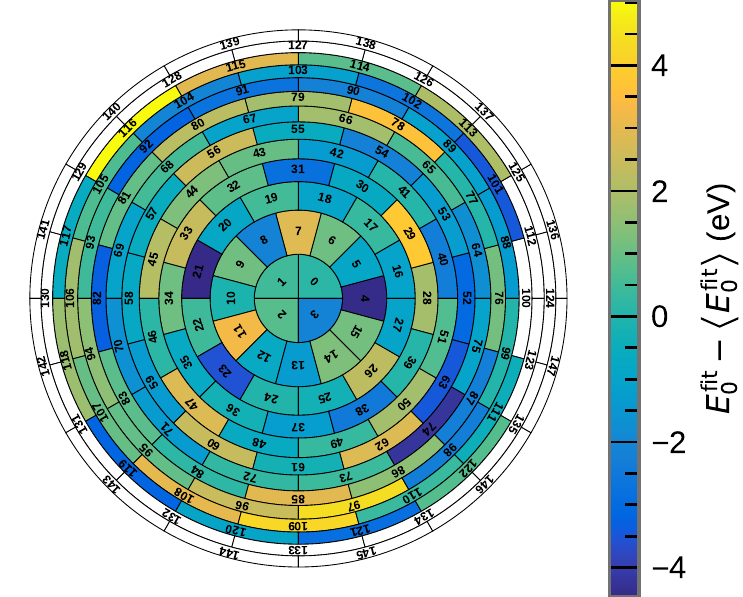}
		\label{Fig:pixel-map}
        \caption{Fitted endpoint $E^{\text{fit}}_0$ for each pixel. All golden scans were stacked and the spectrum of each pixel was fitted in the nominal fit range. Within the uncertainty, no spatial dependence is visible. The white pixels indicate pixels which were excluded from the analysis due to alignment issues or malfunctions, as described in section~\ref{ssec:dataselection}.}
\end{figure}

Combining all golden scans, single-pixel fits were performed resulting in an endpoint $E^{\text{fit}}_0$ for each pixel, as shown in figure~\ref{Fig:pixel-map}. As a result, we find no spatial (i.e. pixel) dependence of $E^{\text{fit}}_0$ beyond the statistical fluctuation. The standard deviation from the mean endpoint is \SI{2.0}{\electronvolt}, which is consistent with statistical fluctuations. This indicates a good description of the analyzing plane electric potential and the absence of a significantly spatially dependent source potential. 

\begin{figure*}[]
\centering
		\includegraphics[width=170mm]{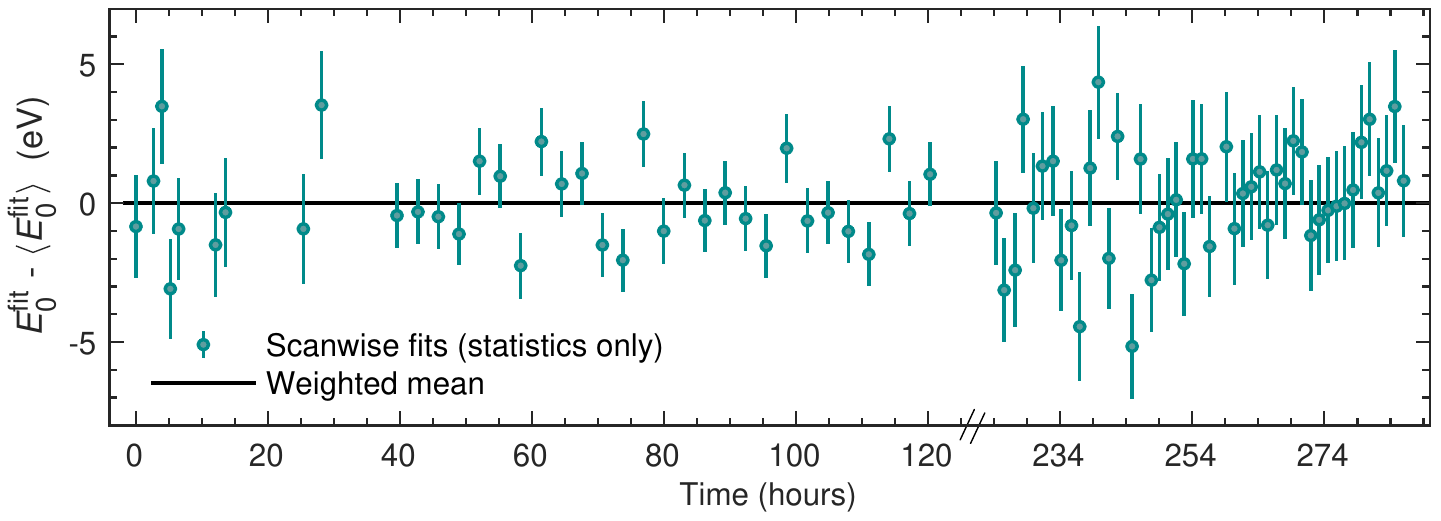}
		\label{Fig:endpoint-evolution}
        \caption{The fitted endpoint for each scan of the golden scan list. For this purpose all pixels were combined into a uniform pixel with an averaged transmission function. Fitting a constant to this endpoint evolution yields a reduced $\chi^2$ of 1.2 and a p-value of 0.14. This demonstrates that the endpoint was stable within statistical fluctuations over the course of almost \SI{300}{\hour} (12.5 days). Note, the scale break at about \SI{125}{\hour} where no $\upbeta$-scans were performed.}
\end{figure*}

Combining all pixels in a uniform fit, we can consider the time evolution of $E^{\text{fit}}_0$, see figure~\ref{Fig:endpoint-evolution}. The data shows excellent stability over the course of 12 days. The standard deviation from the mean endpoint is \SI{1.8}{\electronvolt}, which is again consistent with statistical fluctuations.

\section{Conclusion}
In the First Tritium (FT) measurement campaign, tritium was for the first time circulated through the KATRIN source and first tritium $\upbeta$-electron spectra were recorded. This constitutes a major milestone before the start of the neutrino mass measurement.  

The FT measurements demonstrate the stable operation of the KATRIN source at full column density with \SI{0.5}{\percent} tritium concentration over several days. The beam tube temperature and buffer-vessel pressure could be demonstrated to be stable at the $10^{-5}$ level, which is well below the specified limit. The overall $\upbeta$-decay activity was demonstrated to be stable at the level of $10^{-3}$. 

The first tritium spectra were used to validate and optimize the KATRIN analysis strategy. A selection of distinct techniques for combining data sets and for implementing systematic uncertainties were successfully tested. An excellent agreement of the spectrum calculation with the data was achieved. This agreement is even present for an energy range exceeding the nominal scanning window for neutrino mass measurement by a factor of 10.

The fitted endpoint $E^{\text{fit}}_0$, used as a proxy in this analysis, could be determined with an accuracy of \SI{250}{\milli\electronvolt}. Within this uncertainty, the endpoint did not show any dependence on the fitting range, the column density, or the scanning strategy. Moreover, no radial or azimuthal dependence with regards to the beamline cross-section was observed. Finally, it could be shown that $E^{\text{fit}}_0$ is stable over a time scale of several days. All these properties are essential prerequisites for the neutrino mass measurements.

After this successful commissioning of KATRIN with traces of tritium, the next milestone of KATRIN will be the ramp-up to the nominal source activity and the first neutrino mass campaign which will explore the neutrino mass parameter space at unprecedented sensitivity.

\section*{Acknowledgments}
We acknowledge the support of Helmholtz Association, Ministry for Education and Research BMBF (5A17PDA, 05A17PM3, 05A17PX3, 05A17VK2, and 05A17WO3), Helmholtz Alliance for Astroparticle Physics (HAP), Helmholtz Young Investigator Group (VH-NG-1055),
Max Planck Research Group (MaxPlanck@TUM), and Deutsche Forschungsgemeinschaft DFG (Research Training Groups GRK 1694 and GRK 2149, and Graduate School GSC 1085 - KSETA) in Germany; Ministry of Education, Youth and Sport (CANAM-LM2011019, LTT19005) in the Czech Republic; cooperation with the JINR Dubna (3+3 grants) 2017–2019; and the United States Department of Energy through grants DE-FG02-97ER41020, DE-FG02-94ER40818, DE-SC0004036, DE-FG02-97ER41033, DE-FG02-97ER41041, DE-AC02-05CH11231, DE-SC0011091, and DE-SC0019304, and the National Energy Research Scientific Computing Center. The author S. Mertens gratefully acknowledges the support by the Max Planck Research Group (MPRG) program and especially the MaxPlanck@TUM initiative.

\bibliographystyle{spphys}
\bibliography{Paper.bib}

\end{document}